\documentclass[]{spie}  
\usepackage[]{graphicx}
\usepackage[]{times}
\usepackage[]{subfigure}
\newcommand{\farcs}{\mbox{\ensuremath{.\!\!^{\prime\prime}}}}%
\newcommand{\h}{{\em H}}
\newcommand{\ks}{{\em K}$_{\rm s}$}

\title{Imaging the dense stellar cluster R136 with VLT-MAD}

\author{
M.~A.~Campbell$^{1}$, C.~J.~Evans$^{2}$, J.~Ascenso$^{3}$, A.~J.~Longmore$^{2}$, J.~Kolb$^{4}$, M.~Gieles$^{5}$ \& J.~Alves$^{6}$, 
\normalsize{
\begin{flushleft}
$^{1}$Institute for Astronomy, The University of Edinburgh, Royal Observatory, Blackford Hill, Edinburgh EH9 3HJ, UK;\\
$^{2}$UK Astronomy Technology Centre, Royal Observatory, Blackford Hill, Edinburgh, EH9 3HJ, UK; \\
$^{3}$Centro de Astrof\'{i}sica da Universidade do Porto, Rua das Estrelas, 4150-762 Porto, Portugal; \\
$^{4}$European Southern Observatory, Karl Schwartzschild Str. 2, 85748 Garching bei M\"{u}nchen, Germany;\\
$^{5}$European Southern Observatory, Casilla 19001, Santiago 19, Chile;\\
$^{6}$Calar Alto Observatory-Centro Astron\'{o}mico Hispano-Alem\'{a}n, C/ Jes\'{u}s Durb\'{a}n Rem\'{o}n 2-2, 04004 Almeria, Spain.
\end{flushleft}
}}

\begin{document}
\maketitle

\begin{abstract}

We evaluate the performance of the Multi-conjugate Adaptive optics
Demonstrator (MAD) from \h\/ and \ks\/ imaging of 30~Doradus in the
Large Magellanic Cloud.  Maps of the full-width half maximum (FWHM) of
point sources in the \h\/ and \ks\/ images are presented,
together with maps of the Strehl ratio achieved in the \ks-band
observations.  Each of the three natural guide stars was at the edge
of the MAD field-of-view, and the observations were obtained at
relatively large airmass (1.4-1.6).  Even so, the Strehl ratio
achieved in the second pointing (best-placed compared to the reference
stars) ranged from 15\% to an impressive 30\%.  Preliminary
photometric calibration of the first pointing indicates 5$\sigma$
sensitivities of \ks$\sim$21.75 and \h$\sim$22.25 (from 22 and 12 min
exposures, respectively).

\end{abstract}



\section{INTRODUCTION}\label{intro} 

Correction of atmospheric turbulence with adaptive optics (AO) is an
increasingly important method to improve the performance of large
telescopes.  AO-corrected instruments such as NAOS-CONICA\cite{naco}
now routinely deliver excellent image quality in the near-infrared
when compared to natural seeing.  However, sky coverage has been 
limited by the requirement for a bright natural guide star (NGS)
with which to monitor atmospheric variations.  The introduction of
laser guide stars (LGS) at each of the major 8-m class observatories
has opened-up the night sky for AO-assisted imaging and spectroscopy,
but the field-of-view (FoV) over which the point spread function (PSF)
can be improved remains limited to $\sim$10-20$''$, except in very
favourable conditions.  The concept of multi-conjugate adaptive optics
(MCAO) overcomes this limitation, delivering a larger corrected FoV 
via tomographic reconstruction of multiple turbulent layers using
 multiple (natural or artificial) guide stars\cite{mcao}.

As part of the technology development plan towards the European
Extremely Large Telescope (E-ELT), the Multi-conjugate Adaptive optics
Demonstrator (MAD)\cite{marchetti2,marchetti3,marchetti} was developed
by the European Southern Observatory (ESO) as a visiting instrument
for the Very Large Telescope (VLT) in Chile.  Three Shack-Hartmann
wavefront sensors are used to observe three NGS across a 2$'$ circular
field, thereby allowing mapping of the turbulence.  The turbulence is
then corrected using two deformable mirrors (operating at $\sim$400
Hz), one conjugated to the ground-layer (i.e. 0\,km), the second
conjugated to 8.5\,km above the telescope.

The high-resolution, near-IR camera used with MAD is the CAmera for
Multi Conjugate Adaptive Optics (CAMCAO), designed and built by a
consortium led by the University of Lisbon \cite{camcao1,camcao2}.
CAMCAO operates over the {\em J}, \h, and \ks\/ bands, with critical (2
pixel) sampling of the diffraction-limited PSF at 2.2$\mu$m.  The
detector is a HAWAII2 2k$\times$2k HgCdTe array, with a pixel scale of
0.028$''$/pixel, giving a FoV of 57$''$x57$''$.  A useful feature is
that the camera can be moved within the 2$'$ field without
requiring positional offsets of the telescope, meaning that the AO
loop can remain closed.

MAD was installed at the Nasmyth visitor focus of UT3 Melipal in early
2007 for technical evaluation.  The performance of the instrument was
impressive\cite{marchetti}, with the observations from the first run
already beginning to yield science results\cite{mad1,mad2}.  Here we
present preliminary analysis of MAD \h- and \ks-band imaging of the
central region of 30~Doradus, the most massive star-formation region
in the Large Magellanic Cloud (LMC).

\section{SCIENTIFIC BACKGROUND}

30 Doradus is one of the largest star-forming regions in the Local
Group. At its core is the dense star cluster R136, with stellar ages
in the range of 2-4\,Myr and a total stellar mass of $\sim$5x$10^{4}$
$M_{\odot}$, putting it on a par with massive clusters found in
starburst and interacting galaxies such as M51, M82 and the
Antennae. As a proto-typical `starburst', significant observational
effort has been directed toward the region over the past 20
years. 30~Dor provides an excellent laboratory to study star
formation, and also offers insight into the nature of distant
super-star-clusters and starburst galaxies, for which we only have
integrated properties -- if we can understand large clusters on our
doorstep, we can be more confident of accurate interpretation of those
far away.

However, the formation of large clusters remains poorly understood. In
a relatively short period of time (a few Myrs) a complicated mixture of
various physical processes take place that transform a giant molecular
cloud (GMC) into a star cluster. In the competitive accretion model of
star formation\cite{1}, massive stars form in the centre of the
gravitational well of the GMC. This model successfully explains
observations that show massive stars to be more concentrated in young
clusters such as in the Orion Nebula\cite{2}, yet the situation is
complicated by the fact that massive stars will also tend to move
toward the cluster centre over their lifetimes due to dynamical
interactions, i.e. mass segregation.

To date, ground-based optical imaging and spectroscopy has been used
in 30 Dor to study the initial mass function (IMF), reddening,
star-formation history, stellar content and
kinematics\cite{3,4,5,6,7}. However, only with near-IR imaging are we
able to disentangle multiple objects, and to identify nascent stars
that are still partially embedded in their gas clouds. Indeed, {\em
HST}-NICMOS observations of some of the complex nebular structures in
30~Dor have provided evidence of triggered star-formation, showing the
region to be a two-stage starburst \cite{8,9}. The core of R136 is too
dense for traditional (seeing-limited) ground-based techniques, with 
a core radius of only $\sim$1\farcs3\cite{15}. Only
with the arrival of {\em HST} was R136 resolved in optical and UV
images \cite{10,11,12}, with follow-up spectroscopy revealing a
hitherto unprecedented concentration of the earliest O-type stars
\cite{13}. The conclusion of these studies was that the high- and
intermediate-mass IMF is `completely normal', i.e. Salpeter-like
\cite{13}. This contrasts with results from AO-corrected, near-IR
images from the ESO 3.6-m that, when combined with the {\em HST} data, found
evidence of mass-segregation via a flattening of the IMF in the core
\cite{14}. Unfortunately, by virtue of using such novel technology,
the 3.6-m images were limited to a relatively small field-of-view,
with the core of R136 in one quadrant of a 12\farcs8 x 12\farcs8 field. 

The offer to the community of Science Demonstration (SD) observations
with MAD presented the perfect opportunity to revisit these
conflicting results, while also providing further tests of its
capabilities.  There is NICMOS \h-band imaging of R136 \cite{17}, but
the spatial resolution is lower and the field-of-view is smaller than
that provided by MAD.  Our primary objective with the new MAD
observations is to probe the intermediate-mass IMF of R136 at
unparalleled spatial resolution. For instance, do we see evidence for
a flattening of the intermediate-mass IMF with radius?  

Secondly, from direct star counts the luminosity profile of R136
appears to be best described by two components\cite{15}, with a break
at 10$''$.  There is significant extinction at this radius in the
cluster so, by penetrating the gas and dust more successfully than in
the optical, we should be able to provide empirical constraints on the
outer component of the luminosity profile, investigating the so-called
`excess light' that is predicted to originate from rapid gas removal
in the early stages of cluster evolution\cite{16}.  Determining if R136
is an expanding group or a dynamically-stable star cluster will 
serve as an important ingredient in the recent debate on the importance of
`infant mortality' of young clusters \cite{chandar,gieles}.

\section{OBSERVATIONS \& DATA REDUCTION}

\begin{figure}
\begin{center}
\includegraphics[scale=1.0]{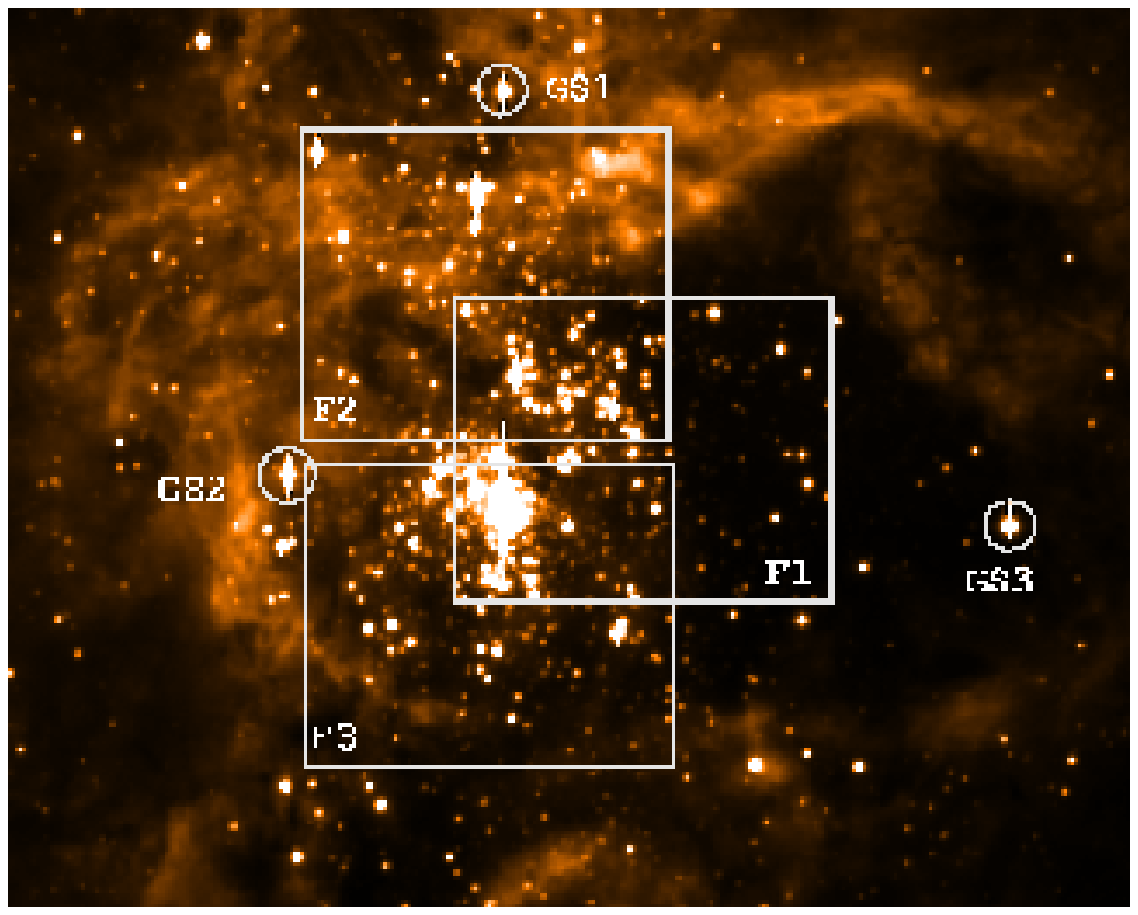}
\caption{$V$-band image of the central part of 30~Dor from the wide-field imager (WFI) 
on the ESO/Max Planck Gesellschaft 2.2-m telescope.  North is towards
the top of the page, east towards the right.  The reference stars used
for the MCAO correction are shown (GS1, 2 \& 3), together with the spatial
extent of the \h-band images of the three observed fields.}\label{obs}
\vspace{0.5cm}
\includegraphics[scale=1.0]{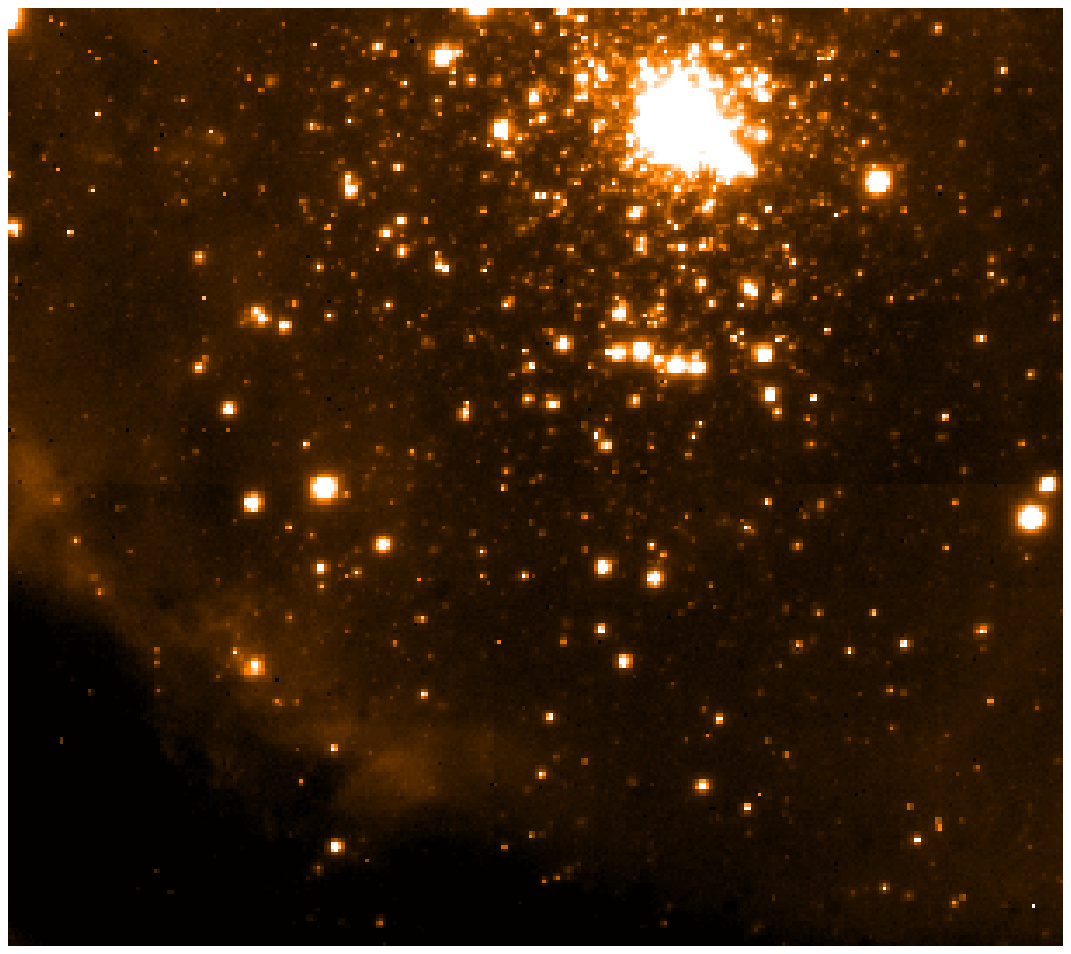}
\caption{Combined VLT-MAD \ks\/ image of Field 3 (same orientation as Figure~\ref{obs}).}\label{K3}
\end{center}
\end{figure}

The data presented here were obtained from one of twelve SD
programmes, observed in November 2007 and January
2008.  \h- and \ks-band images were obtained of three pointings in
the central region of 30~Dor, as shown in Figure~\ref{obs}.  The
central co-ordinates for the CAMCAO observations of Field 1 were
$\alpha =~$05$^{\rm h}$38$^{\rm m}$46.5$^{\rm s}$, $\delta =
-$69$^\circ$05$'$52$''$ (J2000.0).  Fields 2 and 3 were offset from
this first pointing by $-$25$''$ in right ascension and $\pm$25$''$ in
declination.  The three NGS used for wavefront sensing are also shown
in Figure~\ref{obs} -- these are Parker \#952 (GS1), \#499 (GS2), and
\#1788 (GS3), with {\em V}\,=\,12.0, 11.9, and 12.0, respectively\cite{4}.
Closer NGS with a more even spread over the three pointings would have
been preferable, but these were the only three stars available that
were both uncrowded and suitably bright in the $V$ band.  A positive
side-effect of this is that, for the three observed fields, the wider
separation allows tests of the MCAO performance for a variety of
distances/orientations to the NGS.

The observations are summarised in Table~\ref{obsinfo}.  The detector
integration time (DIT) for all of the observations was 2\,s, with 30
integrations (NDIT) for each exposure.  `Batches' of three (\h)
and six (\ks) object and sky frames were interleaved in an
A-B-A-B-A-B-A pattern, yielding total exposures of 12\,min for
each field in the \h\/ band, and 24\,min in \ks.  The science
exposures within each batch were each dithered by 5$''$ -- although
this reduces the effective area of the final combined images, it was
intended to minimise the impact of bad pixels and cosmetic features from
array.  Given the spatial extent of 30~Dor, the sky
offsets were somewhat large ($+$12s of right ascension, $+$13$'$ in
declination) to ensure they were uncontaminated by nebulosity.
Observations were halted mid-way through the \ks\/ exposures for Field
1 on 2008/01/07 due to bad weather, but were completed the following
night.  The airmass of the observations ranged from 1.4 to 1.6.  Note
that the LMC never rises above an altitude of approx. 45$^\circ$ as
viewed from Paranal, i.e. the {\em minimum} zenith distance of the MAD
observations was $\sim$45$^\circ$.  The range of seeing values for
each pointing, as measured by the Differential Image Motion Monitor
(DIMM) at Paranal, are given in Table~\ref{obsinfo}.

\begin{table*}[t]
\begin{center}
\caption{Summary of the VLT-MAD observations in 30~Doradus.  The total exposure times quoted are
for the final combined images.}\label{obsinfo}
\begin{tabular}{lclcccc}
\hline
Pointing & Band & Date & Total Exp.  & DIMM range & Image FWHM & $<$FWHM$>$ \\
& & & [min] & [$''$] & [$''$] & [$''$] \\
\hline
Field 1 &  \ks  & 2008/01/07 \& 08 & 22 & 0.4-1.8 & 0.10-0.13 & 0.11 \\
Field 2 &  \ks  & 2008/01/07           & 24 & 0.5-1.1 & 0.08-0.10 & 0.09 \\
Field 3 &  \ks  & 2007/11/27           & 23 & 0.6-1.0 & 0.10-0.20 & 0.14 \\
\hline
Field 1 & \h & 2008/01/08           & 12 & 0.3-0.6 & 0.10-0.12 & 0.11 \\
Field 2 & \h & 2008/01/08           & 12 & 0.9-1.1 & 0.08-0.11 & 0.09 \\
Field 3 & \h & 2008/01/08           & 11 & 0.6-1.6 & 0.08-0.15 & 0.12 \\
\hline
\end{tabular}
\end{center}
\end{table*}

The MAD data were reduced with standard {\sc iraf} routines, using
calibration frames from the SD runs to correct for the dark current
and to flat-field all of the object and sky exposures.  Median sky
frames were created for each batch of science observations using the
sky frames observed immediately before and/or after, although (in
general) the sky background did not appear to vary strongly over each
sequence of observations.  The sky-subtracted frames were then aligned
with each other and combined.  At this stage we omitted the worst one
or two images in some pointings, hence the exposure times in
Table~\ref{obsinfo} differ slightly to the numbers mentioned
previously.

\begin{figure} [!h]
\begin{center}
\includegraphics[scale=0.8]{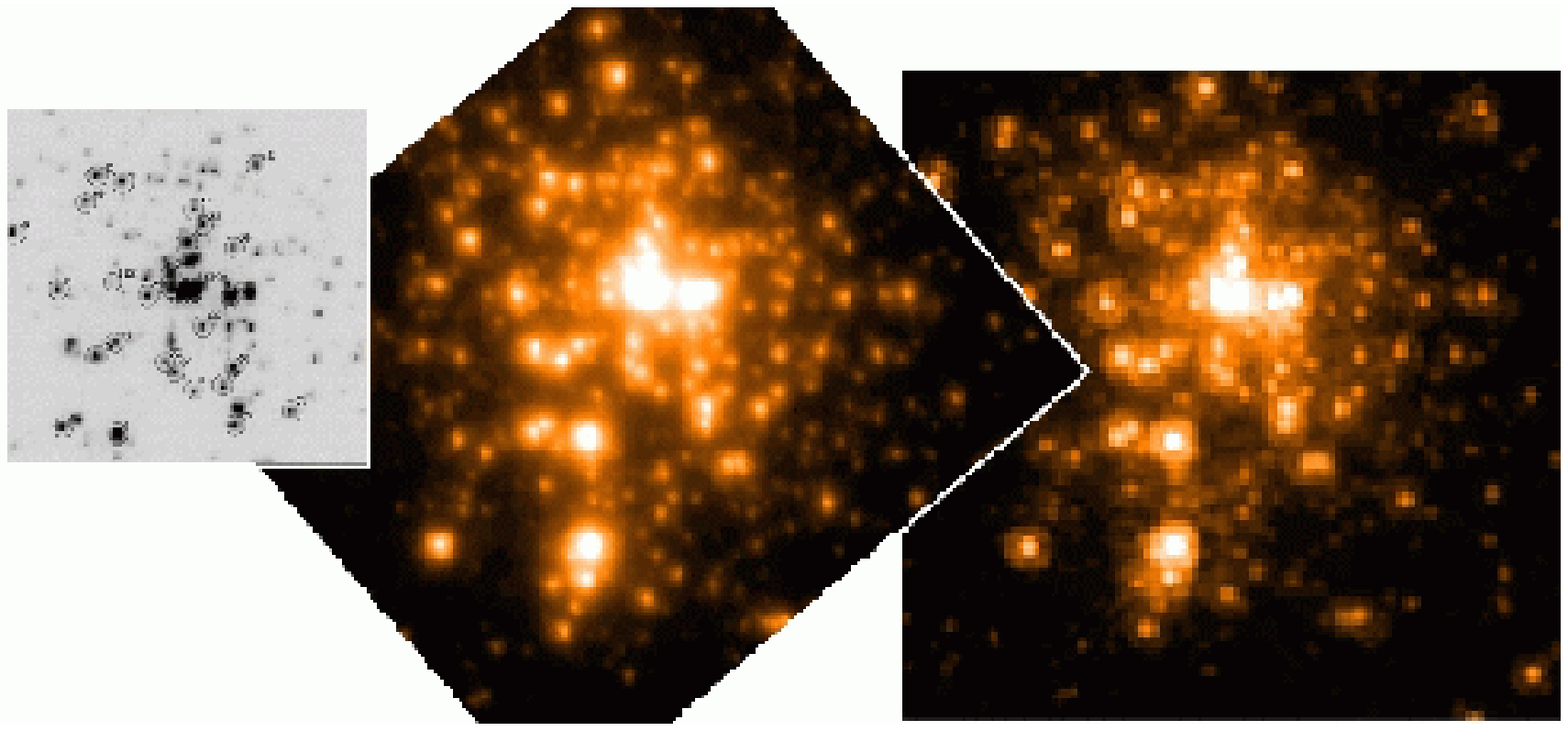}
\caption{Comparison of different images of the core of R136: {\it Left:} {\em HST}-WFPC2 F555W image\cite{13};
{\it Centre:} $\sim$10$''$x10$''$ of the Field 3 \h-band MAD image;
{\it Right:} {\em HST}-NICMOS F160W image\cite{17, 18}.  The MAD and NICMOS intensities are plotted
as log-scales.}\label{montage}
\end{center}
\end{figure}

Although it proved impossible to remove some of detector artefacts 
in the final images (experiments with different $\sigma$-clips led to
unpleasant effects on genuine sources!), these are relatively
low-level features and the final images are truly impressive.
The combined, sky-subtracted \ks-band image for Field 3 is shown in
Figure~\ref{K3}, highlighting the extreme density of stars in the cluster.
In Figure~\ref{montage} we compare the \h-band image of the core
of R136 (from Field 3), with a {\em HST}-WFPC2 optical image\cite{13} (left-hand
image in the montage) and with a F160W (i.e. $\sim${\em H} band) {\em HST}-NICMOS
image\cite{17,18}.  Note that the spatial performance achieved in the MAD image
is comparable to the optical performance of {\em HST}, with improved
spatial performance compared to the near-IR {\em HST} image. Also,
the MAD image does not suffer the diffraction effects seen in the NICMOS
image, making fainter companions easier to identify.

\section{PRELIMINARY ANALYSIS}

To investigate the instrument performance in our three pointings we
first measured the full-width half maximum (FWHM) of 32 stars,
uniformly distributed across each co-added image.  Using stars that
appear to be single objects (that were not saturated and relatively
uncrowded) the position, ellipticity and FWHM were measured, as well
as visual inspection of the radial profiles and contour maps.  These
results are summarised in Table~\ref{obsinfo}.  Residual
ellipiticity/PSF asymmetries are often seen to arise from partial
correction of turbulence\cite{glao}.  Although there is some
variability in the ellipticity results from object-to-object in the
MAD frames, there appears to be no obvious correlation between the
ellipticity of a star and its proximity to the NGS position, nor the
axis of asymmetry and its position in the field.

\begin{figure}
\begin{center}
\includegraphics[scale=0.85]{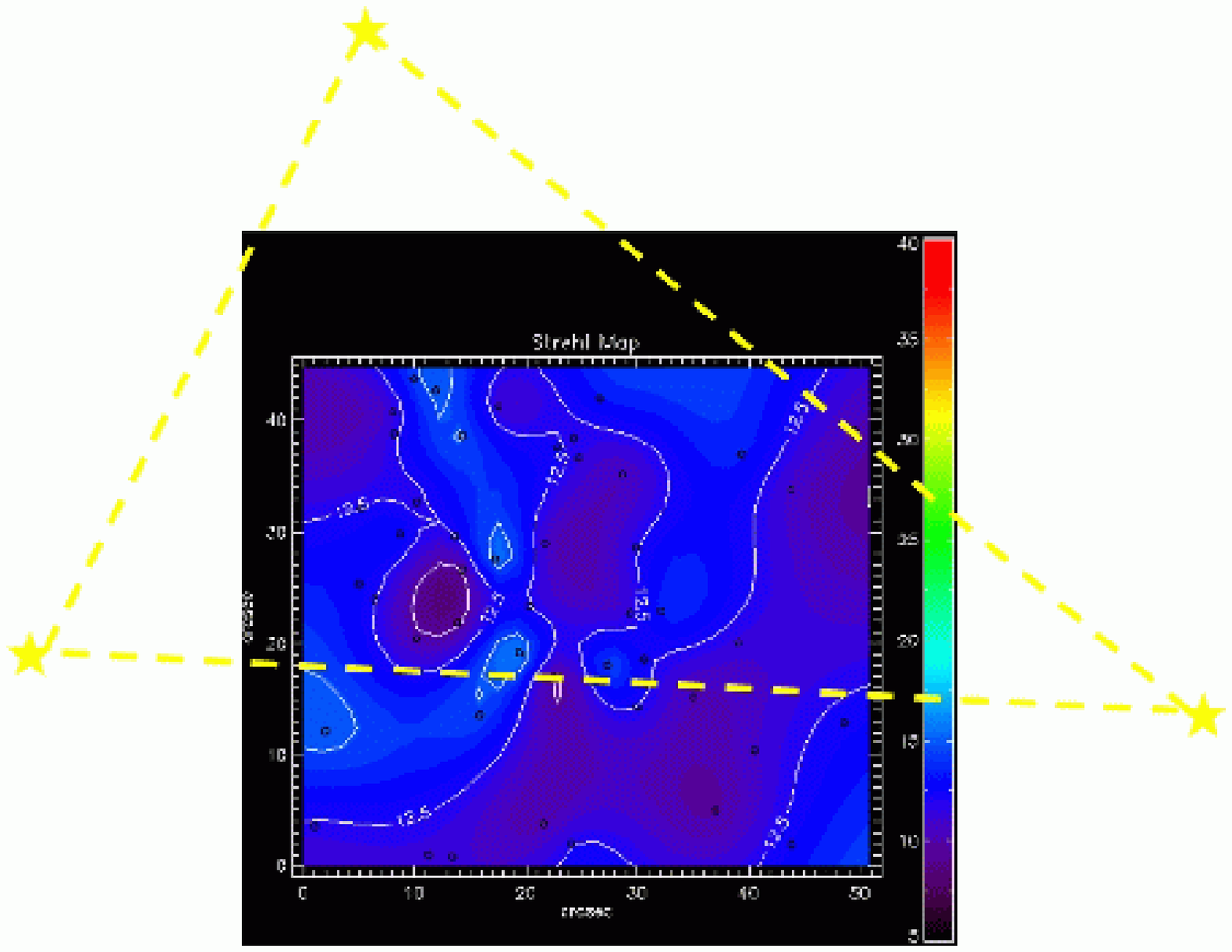}
\caption{\ks-band Strehl (\%) map for Field 1.  The relative positions of the
NGS are indicated by the yellow stars.}\label{strehl_1}
\vspace{1.5cm}
\includegraphics[height=7.65cm]{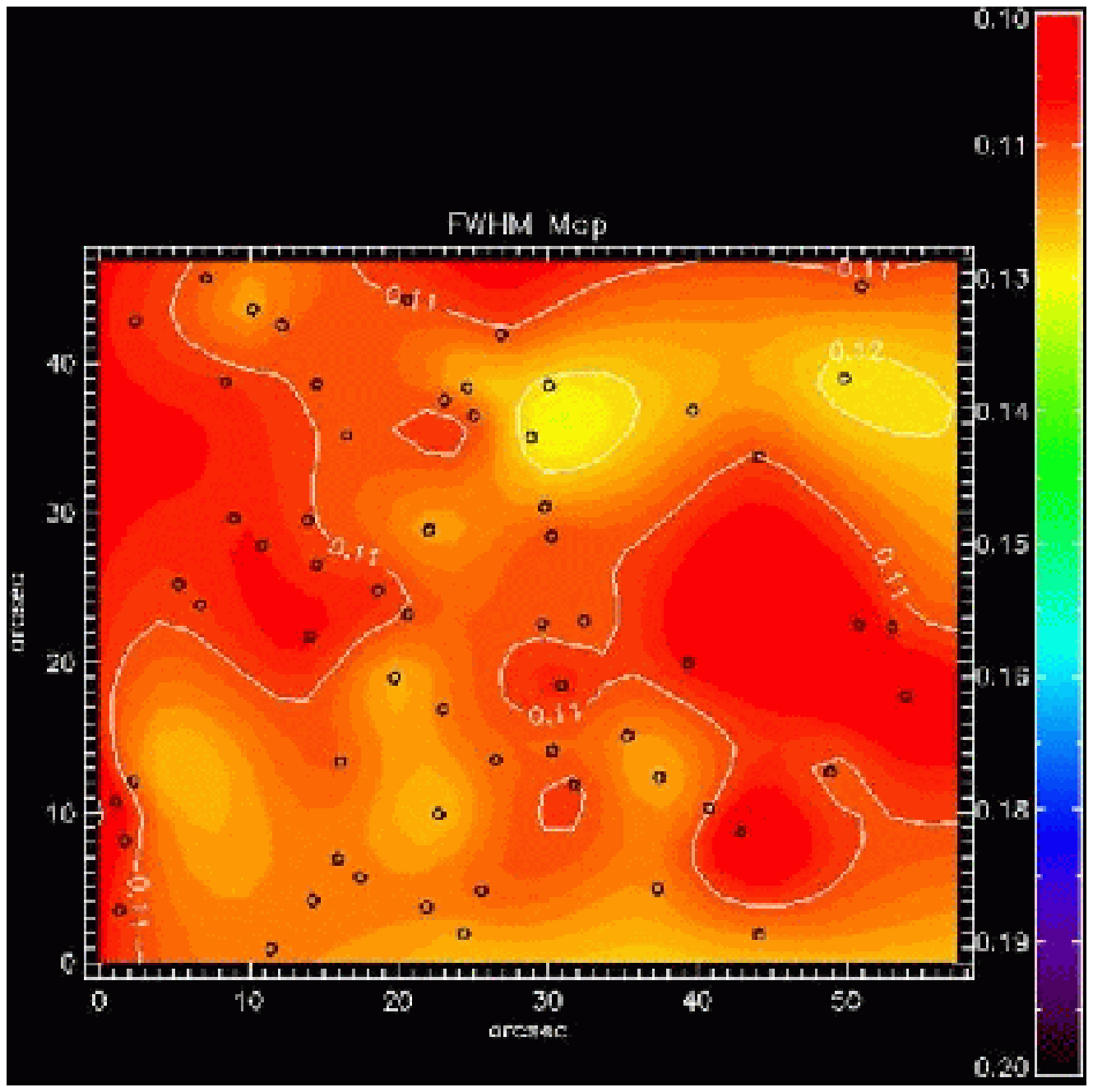}\hspace{0.17cm}\includegraphics[height=7.65cm]{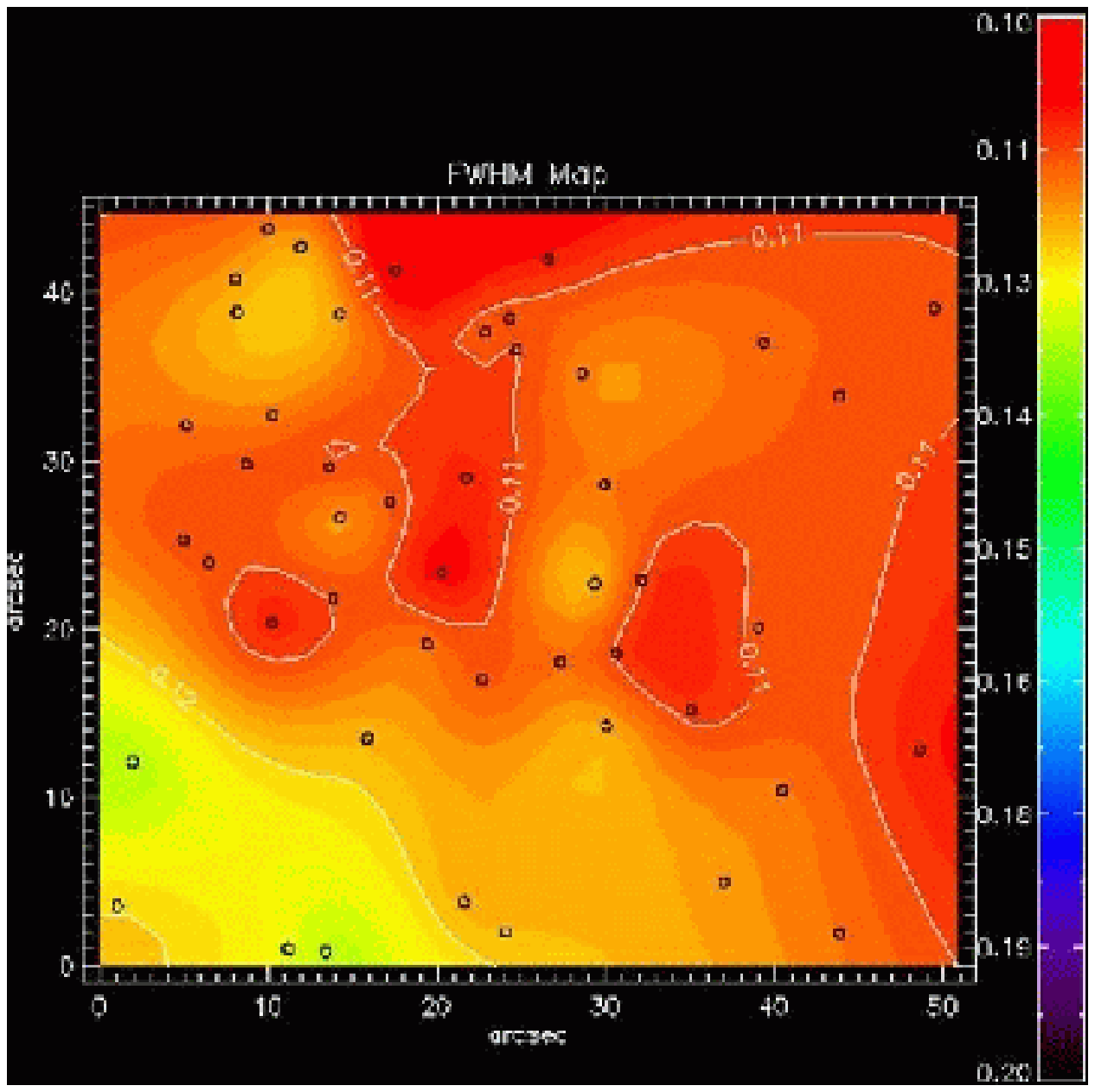}
\caption{\h\/ ({\it left}) and \ks-band ({\it right}) FWHM maps (in arcsec) for Field 1.}\label{fwhm_1}
\end{center}
\end{figure}

\begin{figure}
\begin{center}
\includegraphics[scale=0.85]{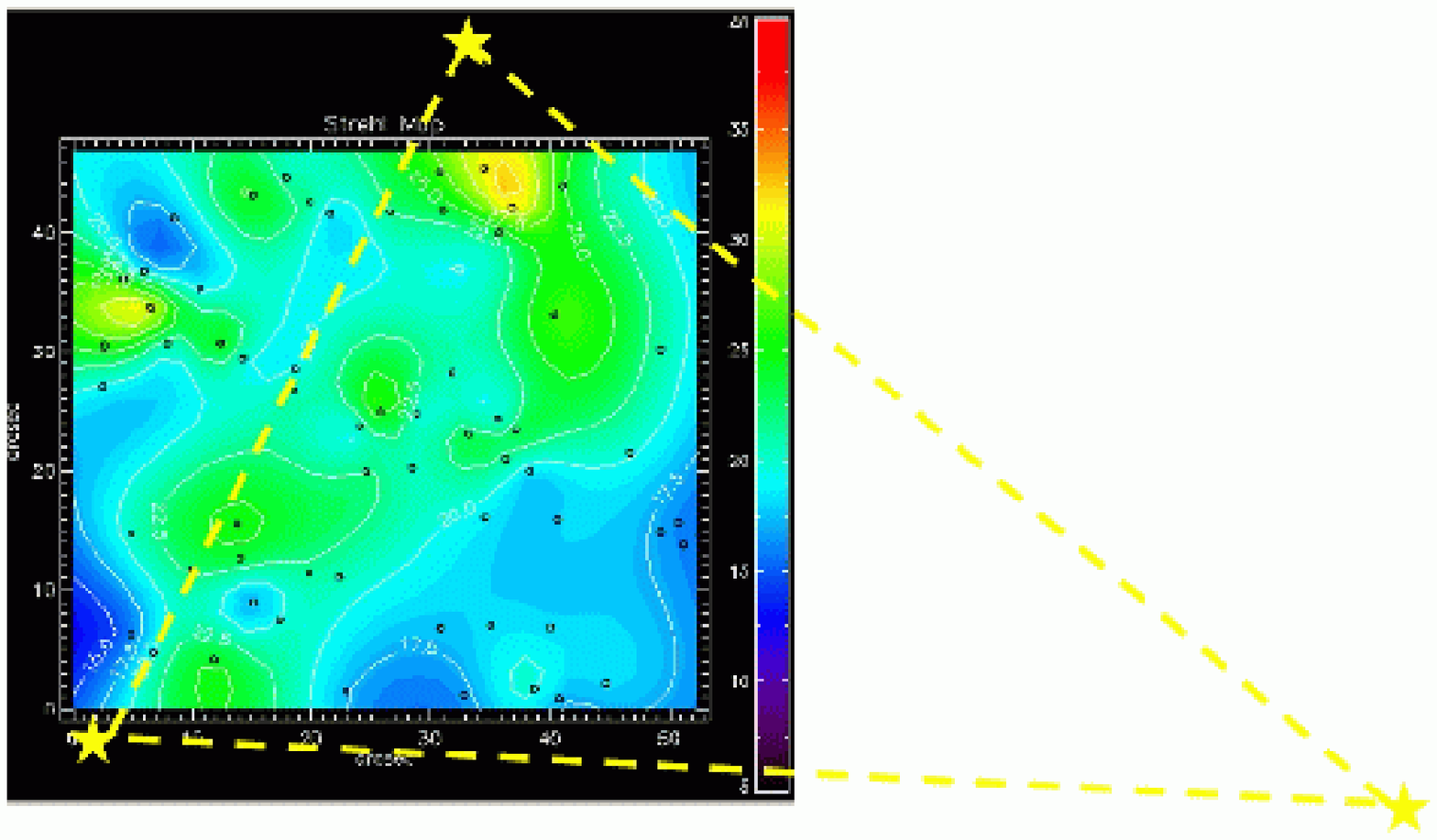}
\caption{\ks-band Strehl (\%) map for Field 2.}\label{strehl_2}
\vspace{1.5cm}
\includegraphics[height=7.65cm]{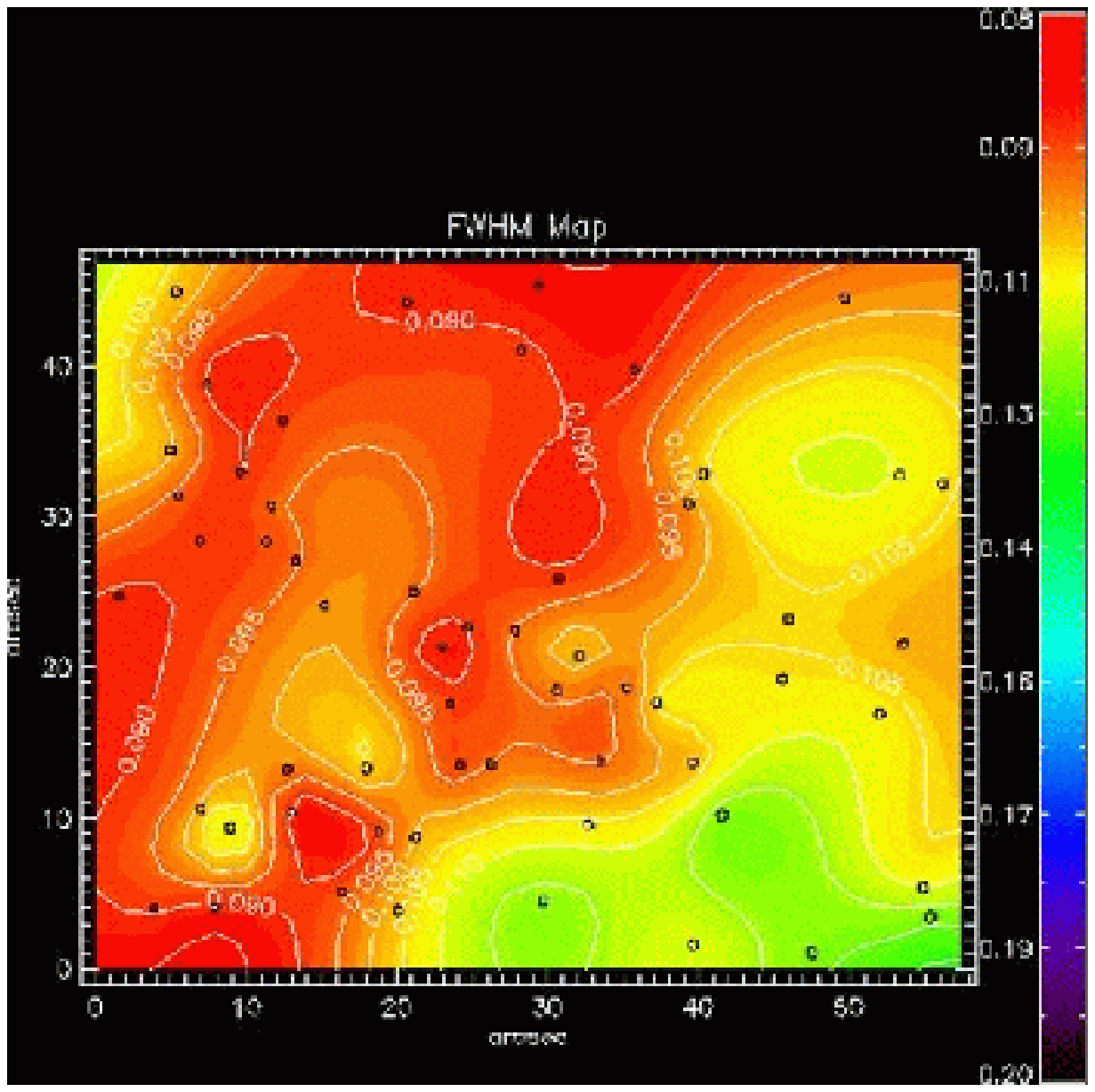}\hspace{0.17cm}\includegraphics[height=7.65cm]{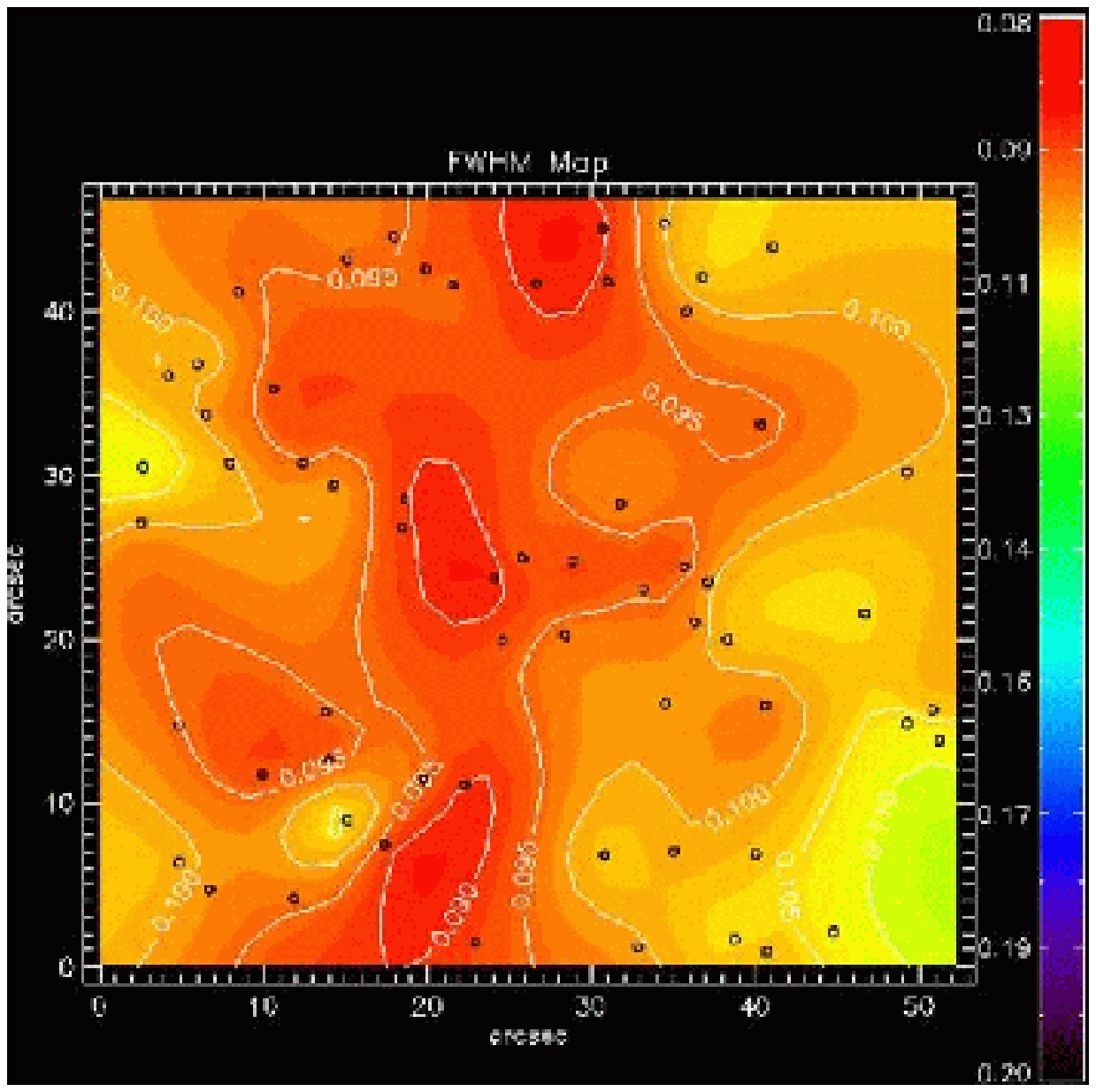}
\caption{\h\/ ({\it left}) and \ks-band ({\it right}) FWHM maps (in arcsec) for Field 2.}\label{fwhm_2}
\end{center}
\end{figure}

\begin{figure}
\begin{center}
\includegraphics[scale=0.85]{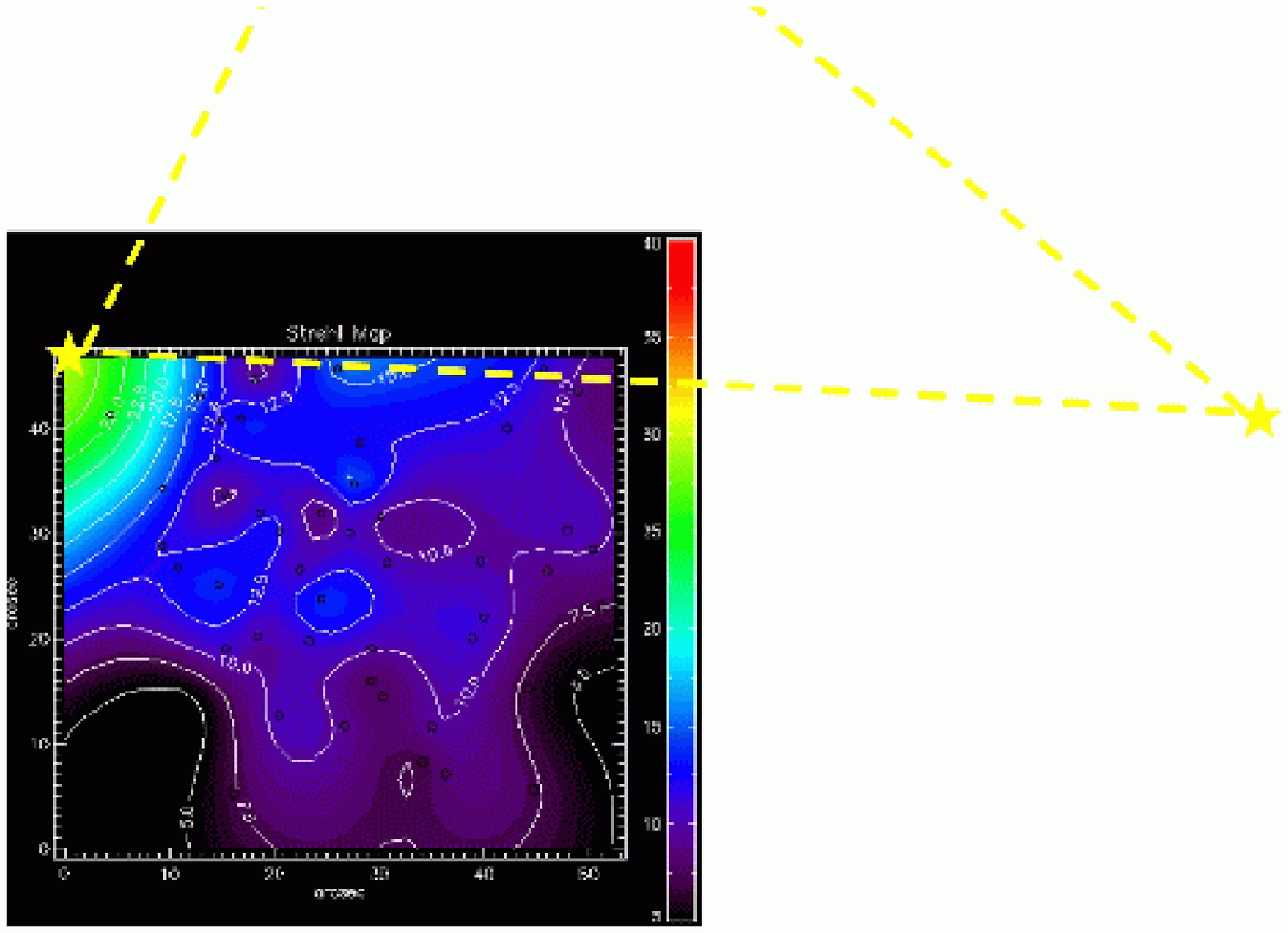}
\caption{\ks-band Strehl (\%) map for Field 3.}\label{strehl_3}
\vspace{1.5cm}
\includegraphics[height=7.65cm]{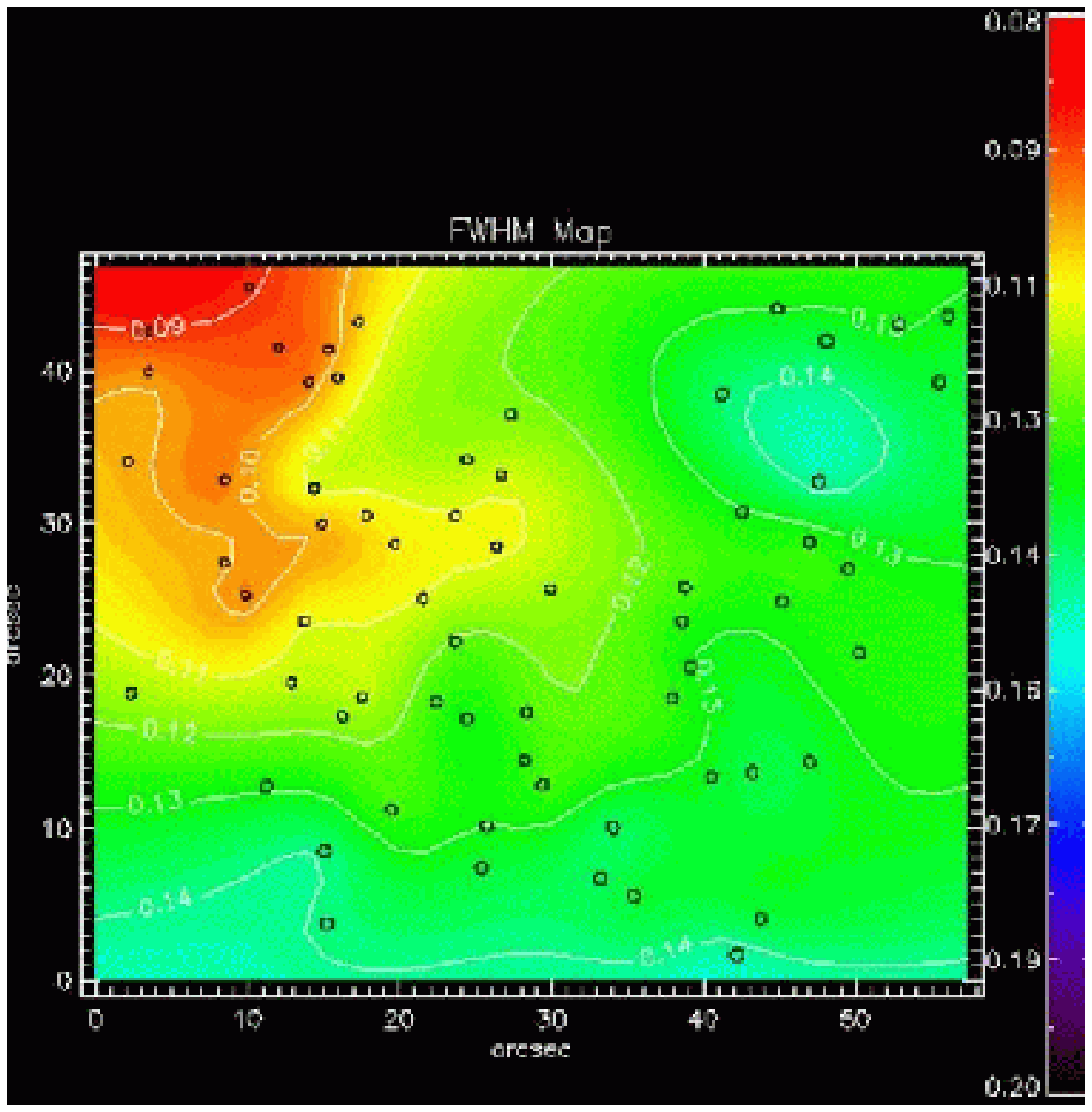}\hspace{0.17cm}\includegraphics[height=7.65cm]{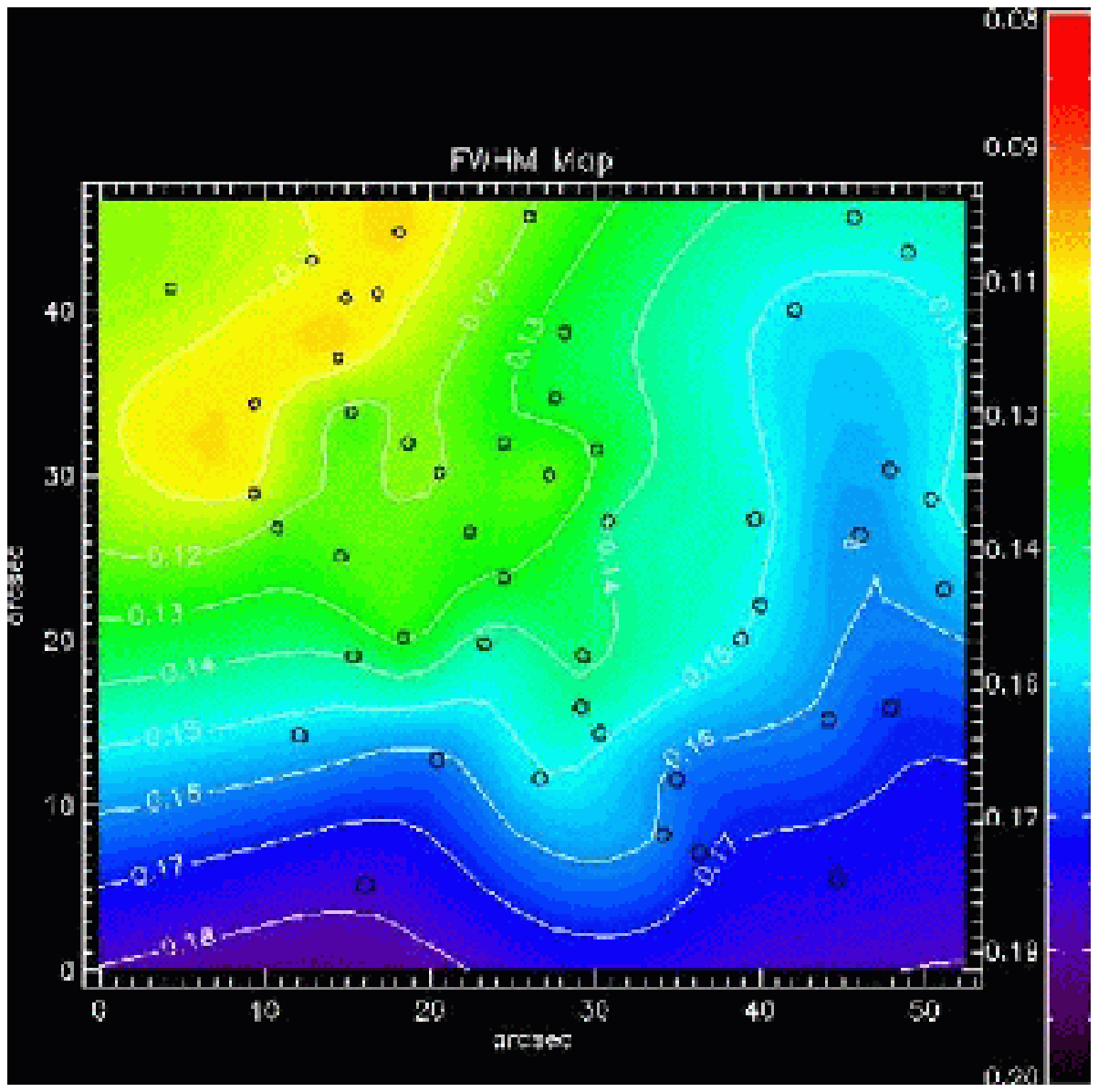}
\caption{\h\/ ({\it left}) and \ks-band ({\it right}) FWHM maps (in arcsec) for Field 3.}\label{fwhm_3}
\end{center}
\end{figure}

A set of IDL routines was used to create maps of the FWHM across
each field, and maps of the Strehl ratio in the \ks-band images --
note that it is not possible to create similar maps for the \h-band
images due to the undersampling of the detector.  The Strehl map for
the \ks-band image of Field 1 is shown in Figure~\ref{strehl_1}
(including the relative positions of the NGS), with the FHWM maps for
both \h\/ and \ks\/ shown in Figure~\ref{fwhm_1}.  Similar maps are
shown in Figures~\ref{strehl_2} and \ref{fwhm_2} for Field 2, and in
Figures~\ref{strehl_3} and \ref{fwhm_3} for Field 3.

The best Strehl ratio achieved in Fields 2 and 3 was 25-30\%, in the
regions closest to the NGS (as one would expect).  The performance in
Field 2 is particularly good, with an average FWHM of 0.09$''$ in both
the \h\/ and \ks\/ images (compared to diffraction limits [$\lambda$/D]
of approx 0.04$''$ and 0.06$''$, respectively).  
Although the Strehl is lower in Field 1, this was in the best position
with respect to all three NGS -- note the uniformity across the
$\sim$50$''$x50$''$ field (Figure~\ref{strehl_1}). This is in strong
contrast to Field 3 (Figure~\ref{strehl_3}), in which the performance
is very good in one corner, but then steeply declines away from the
NGS -- more in keeping with `classical' AO observations.
In general, the performances are comparable with those found from the first round
of MAD observations\cite{marchetti, mad1}.  The Strehl achieved in
Field 2 is comparable to that obtained in observations of the
Trapezium cluster from the first MAD observations\cite{mad1}.  These
were also observed at relatively large airmass (1.4 to 1.7), but had
the advantage of two NGS within the CAMCAO field, with the third just
outside it.

Photometric analysis and calibration of the 30~Dor data is now
underway.  Given the large density of stars in the images we are using
PSF-fitting methods in {\sc daophot} -- the observed PSFs are slightly
variable across the field so the model PSF is also allowed to vary.
For a first estimate of sensitivities we have used the Two Two Micron
All Sky Survey (2MASS) catalogue to calibrate the images of Field~1.
In such a dense region there are a limited number of well-resolved
2MASS targets, with 14 in the \h\/ band with quality ratings of A, B or
C, and 10 in the \ks\/ image (which has a smaller effective area owing
to a larger number of dithers).  The zero points were ZP$_{K}$\,=\,$-$26.30$
\,\pm\,$0.35 and ZP$_{H}$\,=\,$-$26.89$\,\pm\,$0.30.  Assuming, for now, that
the MAD filters match those of 2MASS, only zero point corrections
have been applied to the observed magnitudes -- this gives 5-$\sigma$
sensitivity estimates of \ks\/ $\sim$21.75 and \h\/ $\sim$22.25.  Work is now
proceeding on photometric calibrations using published photometry
from higher-resolution imaging than 2MASS.

\section{SUMMARY}

The observations presented here provide a further demonstration of 
the excellent performance of VLT-MAD. The image quality (as measured by the FWHM
of the PSF) across Fields 1 and 2 was uniform to within 0\farcs03.
Such stability across `wide' fields (compared to classical AO
observations) is very encouraging in advance of future multi-laser,
MCAO systems on 8-m class telescopes, and the developement of
`wide-field', AO-corrected, imagers and spectrometers for the E-ELT.

MAD has also provided us with a truly unique view
of the stellar populations in 30~Doradus, in terms of both spatial
resolution and spatial coverage compared to previous studies.  Work is
now underway on the photometric calibrations and the resulting
catalogue will be used to investigate the IMF and star-formation
history of this important cluster.

\acknowledgments      
We are grateful to the ESO staff, in particular to Paola Amico at
Paranal and Enrico Marchetti in Garching, for their advice and
suggestions.  We thank Morten Andersen for his NICMOS image in advance
of publication, also Nate Bastian, Bernhard Brandl, Jacco van Loon and
Hans Zinnecker for useful input and discussion.
\
\
\bibliography{mad}   

\begin{thebibliography}{10}

\bibitem{naco}
{Rousset}, G.~{\rm et al}. {\em \,SPIE,}~{\bf 4839},  140 (2003).

\bibitem{mcao}
{Raggazoni}, R.~{\rm et al}. {\em \,Nature,}~{\bf 403},  54 (2000).

\bibitem{marchetti2}
{Marchetti}, E.~{\rm et al}. {\em \,SPIE,}~{\bf 4839},  317 (2003).

\bibitem{marchetti3}
{Marchetti}, E.~{\rm et al}. {\em \,SPIE,}~{\bf 5490},  236 (2004).

\bibitem{marchetti}
{Marchetti}, E.~{\rm et al}. {\em \,The Messenger,}~{\bf 129},  8 (2007).

\bibitem{camcao1}
{Amorim}, A.~{\rm et al}. {\em \,SPIE,}~{\bf 5492},  1699 (2004).

\bibitem{camcao2}
{Amorim}, A.~{\rm et al}. {\em \,SPIE,}~{\bf 6269},  164 (2006).

\bibitem{mad1}
{Bouy}, H.~{\rm et al}. {\em \,A\&A,}~{\bf 477},  681 (2008).

\bibitem{mad2}
{Gullieuszik}, M.~{\rm et al}. {\em \,A\&A,}~{\bf 483},  L5 (2008).

\bibitem{1}
{Bonnell}, I. A. \&~{Bate}, M.~R. {\em \,MNRAS,}~{\bf 370},  488 (2006).

\bibitem{2}
{Hillenbrand}, L. A. \&~{Hartmann}, L.~W. {\em \,ApJ,}~{\bf 492},  540 (1998).

\bibitem{3}
{Melnick}, J. {\em \,A\&A,}~{\bf 153},  235 (1985).

\bibitem{4}
{Parker}, J.~W. {\em \,AJ,}~{\bf 106},  560 (1993).

\bibitem{5}
{Bosch}, G.~{\rm et al}. {\em \,A\&A,}~{\bf 380},  137 (2001).

\bibitem{6}
{Parker} J. W. \&~{Garmany}, C.~D. {\em \,AJ,}~{\bf 106},  1471 (1993).

\bibitem{7}
{Selman}, F. {\em \,A\&A,}~{\bf 341},  98 (1998).

\bibitem{8}
{Walborn}, N. R.~{\rm et al}. {\em AJ}~{\bf 117},  225 (1999).

\bibitem{9}
{Walborn}, N. R.~{\rm et al}. {\em \,AJ}~{\bf 124},  1601 (2002).

\bibitem{15}
{Mackey}, A. D. \&~{Gilmore}, G.~F. {\em \,MNRAS}~{\bf 338},  85 (2003).

\bibitem{10}
{Hunter}, D. A.~{\rm et al}. {\em \,ApJ}~{\bf 448},  179 (1995).

\bibitem{11}
{Hunter}, D. A.~{\rm et al}. {\em \,ApJ}~{\bf 113},  1691 (1997).

\bibitem{12}
{Sirianni}, M.~{\rm et al}. {\em \,ApJ}~{\bf 533},  203 (2000).

\bibitem{13}
{Massey}, P. \&~{Hunter}, D.~A. {\em \,ApJ}~{\bf 493},  180 (1998).

\bibitem{14}
{Brandl}, B.~{\rm et al}. {\em \,ApJ}~{\bf 466},  254 (1996).

\bibitem{17}
{Andersen}, M. {\em PhD Thesis, Potsdam University}  (2004).

\bibitem{16}
{Bastian}, N. \&~{Goodwin}, S.~P. {\em \,MNRAS}~{\bf 369},  L9 (2006).

\bibitem{chandar}
{Chandar}, R.~{\rm et al}. {\em ApJ}~{\bf 650},  L111 (2006).

\bibitem{gieles}
{Gieles}, M.~{\rm et al}. {\em ApJ}~{\bf 668},  268 (2007).

\bibitem{18}
{Andersen}, M.~{\rm et al}. {\em in press}  (2008).

\bibitem{glao}
{Tokovinin}, A. {\em PASP}~{\bf 116},  941 (2004).

\end{thebibliography}
\bibliographystyle{spiebib}   

\end{document}